\documentclass{article}

\usepackage{float}
\usepackage{PRIMEarxiv}
\usepackage{enumitem}
\usepackage[utf8]{inputenc} % allow utf-8 input
\usepackage[T1]{fontenc}    % use 8-bit T1 fonts
\usepackage{hyperref}       % hyperlinks
\usepackage{url}            % simple URL typesetting
\usepackage{booktabs}       % professional-quality tables
\usepackage{amsfonts}       % blackboard math symbols
\usepackage{nicefrac}       % compact symbols for 1/2, etc.
\usepackage{microtype}      % microtypography
\usepackage{lipsum}
\usepackage{fancyhdr}       % header
\usepackage{graphicx}       % graphics
\title{An Artificial Intelligence Life Cycle: From Conception to Production
\author{Daswin De Silva and Damminda Alahakoon \\
 Centre for Data Analytics and Cognition (CDAC)\\
 La Trobe University, Victoria, Australia\\
  \texttt{d.desilva@latrobe.edu.au} \\}

}
\begin{document}
\maketitle
\let\thefootnote\relax\footnotetext{D. De Silva and D. Alahakoon, "An Artificial Intelligence Life Cycle: From Conception to Production," arXiv Preprint, 2021.}
\begin{abstract}
Drawing on our experience of more than a decade of AI in academic research, technology development, industry engagement, postgraduate teaching, doctoral supervision and organisational consultancy, we present the 'CDAC AI Life Cycle', a comprehensive life cycle for the design, development and deployment of Artificial Intelligence (AI) systems and solutions. It consists of three phases, Design, Develop and Deploy, and 17 constituent stages across the three phases from conception to production of any AI initiative. The 'Design' phase highlights the importance of contextualising a problem description by reviewing public domain and service-based literature on state-of-the-art AI applications, algorithms, pre-trained models and equally importantly ethics guidelines and frameworks, which then informs the data, or Big Data, acquisition and preparation. The 'Develop' phase is technique-oriented, as it transforms data and algorithms into AI models that are benchmarked, evaluated and explained. The 'Deploy' phase evaluates computational performance, which then apprises pipelines for model operationalisation, culminating in the hyperautomation of a process or system as a complete AI solution, that is continuously monitored and evaluated to inform the next iteration of the life cycle. An ontological mapping of AI algorithms to applications, followed by an organisational context for the AI life cycle are further contributions of this article. 
\newline
\end{abstract}

% keywords can be removed

\section{Introduction}
Artificial Intelligence (AI) is transforming the constitution of human society; the way we live and work. AI is enabling new opportunities for strategic, tactical and operational value creation in all industry sectors and disciplines, including commerce, sciences, engineering, humanities, arts, and law \cite{brynjolfsson2017can}. Private investment in AI was US\$70 billion in 2019, of which start-ups were \$37 billion, leading up to worldwide revenues of approximately \$150 billion in 2020, and this is likely to surpass \$300 billion in 2024 \cite{savage2020race}. In parallel to commercial interests, AI must be leveraged for social impact with initiatives such as AI4SG \cite{tomavsev2020ai} focusing on interdisciplinary partnerships for AI applications that  achieve the United Nations’ Sustainable Development Goals (SDG) \cite{sdgs2015united} and promote public interest technologies for social welfare \cite{mcguinness2021power}. Globally, the signs of an AI race are evident, as the world's superpowers contend for leadership in AI research, investment, patents, products and practice \cite{savage2020race, fleming2020world, polcumpally2021artificial}, as well as the contentious pursuit of an Artificial General Intelligence\cite{naude2020race, wang2007introduction}. Autonomous weapons, social exclusion, microtargeted disinformation, excess energy consumption, underpaid click-work, and technological unemployment are complex and imminent challenges of AI that require global collaboration, not competition. Fortunately, academia, industry and governments alike are affording an equal importance to AI, data and robot ethics, such as the lethal autonomous weapons pledge \cite{lawp2017}, IEEE Ethically Aligned Design (EAD) \cite{ieeeEAD} and the Ethics Guidelines for Trustworthy AI by the High-Level Expert Group on AI (HLEG) \cite{AIHLEG}. Despite these rapid advances, we discovered that existing work on AI life cycles and methodologies either do not render comprehensive coverage from conception to production, or are limited in the level of detail of each individual phase.

Drawing on this context of increasing impact, influence and thereby importance of AI, across national, social, economical and personal interests, we present the CDAC AI life cycle that characterises the design, development and deployment of AI systems and solutions. CDAC is the acronym of our research centre, Centre for Data Analytics and Cognition. This life cycle is informed by our experience and expertise at CDAC of more than a decade of AI, across academic research, technology development, industry engagement, postgraduate teaching, doctoral supervision and organisational consultancy. A few highlights from our recent work are, Bunji, an empathic chatbot for mental health support \cite{bunji2021}, solar nowcasting for optimal renewable energy generation \cite{solar2021}, emotions of COVID-19 from self-reported information \cite{adikari2021emotions}, machine learning for online cancer support \cite{adikari2020can, de2018machine}, self-building AI for smart cities \cite{alahakoon2020self}, an incremental learning platform for smart traffic management \cite{nallaperuma2019online}  and a reference architecture for industrial applications of AI \cite{de2020toward}. We anticipate our contribution will create awareness, instil knowledge, stimulate discussion and debate that will inform research, applications and policy developments of AI for humanity. 

%share this knowledge in the public interest and anticipate it will 
\section{The CDAC AI Life Cycle}

\begin{figure} [h]
  \centering
  \includegraphics[width=1.0\textwidth]{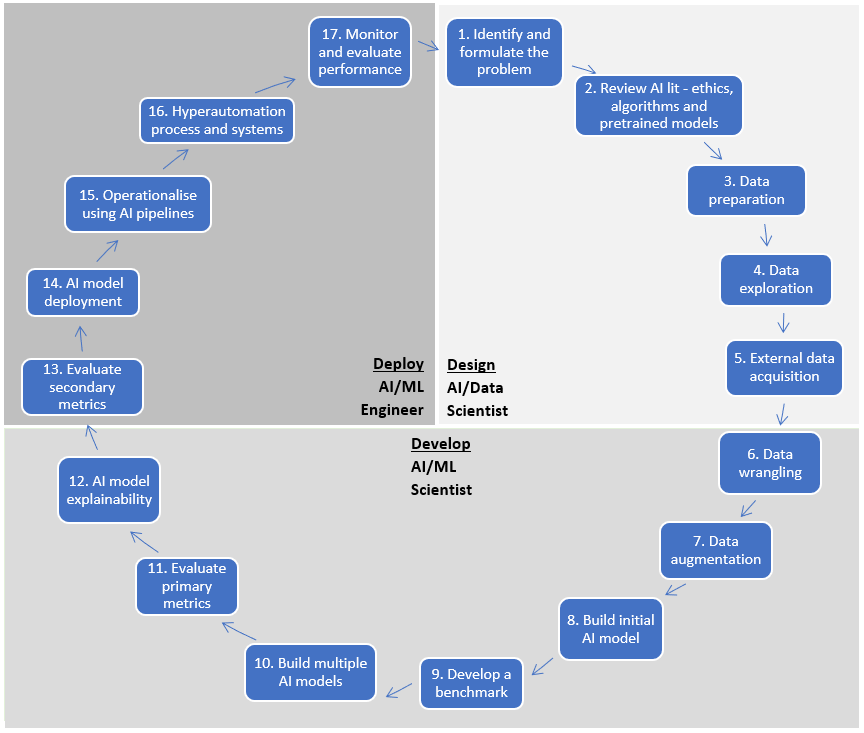}
   \caption{The CDAC AI Life Cycle}
  \label{fig:fig1}
\end{figure}

Figure 1 illustrates the complete AI Life Cycle, where the shaded parallelograms represent the three phases and corresponding human expertise: 1) Design - AI/data scientist, 2) Develop - AI/ML scientist and 3) Deploy - AI/ML engineer. The size of the AI team corresponds to the size of the project, but a minimum of one AI scientist and one AI engineer is recommended. In the following subsections, the 17 stages are prescribed as lists of activities that must be considered and undertaken for successful completion of that stage.

\subsection{Identify and formulate the problem}
\begin{itemize} 
   \item Define the problem in terms of environment, entities and data 
   \item How is the problem currently solved? 
   \item What is the flow of the solution (steps/phases)? 
   \item What are the alternate solutions and flows?
   \item Where/how are the rules defined for the solution?
   \item What data is collected, how is it stored, what historical data is available? 
\end{itemize}

\subsection{Review AI literature - ethics, algorithms and pre-trained models}
\begin{itemize} 
   \item Knowing the problem and its context, review applicable ethics guidelines, ethics frameworks, the state-of-the-art in AI algorithms, models and pre-trained models that have been used to solve this or similar problems 
   \item Pre-trained models like AlexNet \cite{krizhevsky2012imagenet} , ResNet \cite{he2016deep}, BERT \cite{devlin2018bert} and GPT  \cite{radford2018improving} are trained on large (billion+) datasets and they can be re-purposed, retrained or fine-tuned instead of building from scratch 
   \item Where to look: Google Scholar, research labs (e.g. CDAC)  publishing platforms (e.g. Medium), Q\&A sites (e.g. Stack Exchange), code repositories (e.g. GitHub), cloud platforms (e.g. Azure) and Social media (e.g. Twitter) 
   \item What to look for: Literature reviews, commentaries, letters, articles, op-ed, Case studies, best practices, product/tool documentation, tutorials, demonstrations, API documentation, forum responses, upvotes and likes  
\end{itemize}

\subsection{Data preparation}
\begin{itemize} 
   \item All data sources as a Single Version of Truth (SVOT) - data as digital representations of the problem description and potential AI solution
      \item A unified data warehouse, data lake or data lakehouse setup
      \item Identify, record and review: data access, data ownership, data stewardship, metadata, data ethics, governance and regulations
\end{itemize}

\subsection{Data exploration}
\begin{itemize} 
   \item Mapping the granularity and inter-relationships of the data
   \item Checking and validating data quality
   \item Compare industry benchmarks and algorithmic baselines for similar problems
\end{itemize}
\subsection{External data acquisition}
\begin{itemize} 
   \item  Short-term acquisition from data brokers and vendors
   \item Public records, credit scoring services, social media content, data sharing agreements
   \item Review ethics guidelines and regulatory requirements
   \item Long-term acquisition from direct stakeholders, by building trust and transparency  
\end{itemize}
\subsection{Data wrangling}
\begin{itemize} 
   \item Transformations: data types, normalisation, relationships, hierarchies 
      \item Values: duplicates, missing, erroneous
      \item Formatting: dates, representations
\end{itemize}
\subsection{Data augmentation}
\begin{itemize} 
   \item Addressing class imbalance: oversampling or undersampling 
      \item Transfer learning from synthetic data
      \item Feature engineering: dynamic time warping
      \item Feature representation: vector symbolic architectures
\end{itemize}
\subsection{Build initial AI model}
\begin{itemize} 
   \item Identify the AI capabilities that correspond to the AI application (or problem domain), and map these on to AI algorithms (See Figure 2 for an example of this mapping)
      \item Select parameters, architecture that represents the nuances of the problem domain
      \item Build the initial AI model - train (learn), validate, test
      \item If this is a pre-trained model then conduct transfer learning or fine tune with zero-one shot learning
\end{itemize}
\subsection{Develop a benchmark}
\begin{itemize} 
   \item Develop a performance benchmark based on the initial model 
      \item Use a common-sense heuristic such as human expertise in solving the same problem
      \item Use this benchmark to evaluate the algorithm/model, as well as the dataset, what does not fit to the model may not be the best representation of the problem 
\end{itemize}
\subsection{Build multiple AI models}
\begin{itemize} 
   \item New AI models should focus on what the initial model fails to capture 
   \item Consider parameter tuning, regularization techniques, data pre-processing for version of the initial model and new models
      \item Gradually increase model complexity, and not vice versa, as this affords model explainability 
      \item The inverse approach is to increase model complexity up to overfitting and then use regularization techniques to generalise 
\end{itemize}
\subsection{Evaluate primary metrics}
\begin{itemize} 
   \item Determine the correct model evaluation metrics
   \item Use literature, API documentation and related case studies to find these metrics, which can range from accuracy, loss, error to precision, recall, convergence etc. 
   \item Understand the metric in terms of, $Outcome = model + error$
   \item An effective evaluation metric should be accurate, robust, agnostic, scalable, interpretable 
      \item Compare all models across default and fine-tuned parameter settings  
      \item Investigate bias variance trade-off across all models
\end{itemize}
\subsection{AI model explainability (or XAI)}
\begin{itemize} 
   \item Apply intrinsic methods (endemic methods to the algorithm) for interpreting and explaining the model output
      \item Apply extrinsic methods for interpretion and explainability 
      \item Extrinsic methods include Partial Dependence Plots (PDP), Individual Conditional Expectation (ICE), Local Interpretable Model Explanation (LIME) and Shapley Addictive Explanations (SHAP)
      \item Also consider newer versions or extensions of algorithms that provide XAI by design 
\end{itemize}
\subsection{Evaluate secondary metrics}
\begin{itemize} 
   \item In addition to intelligence performance, the models should be be computationally effective so that it can be deployed or operationalised. 
      \item Determine CPU usage and memory usage metrics
      \item Determine complexity and convergence metrics 
      \item Consider model compression and deployment options on-prem, device vs cloud
\end{itemize}
\subsection{AI model deployment}
\begin{itemize} 
   \item Also known as model serving, model scoring, production
   \item Review and identify the most time-efficient and adaptable deployment approach
      \item Consider Real-time vs batch execution/prediction
      \item Consider number of end-users, applications
      \item What are the expected formats of output?
      \item What is the expected turnaround time?
      \item what is the frequency of use?
\end{itemize}

\subsection{Operationalise using AI pipelines (MLOps, AIOps)}
\begin{itemize} 
    \item Moving from standard deployment to AI pipelines using containers and microservices 
    \item A microservice is an application with an atomic function (e.g. classifier or regressor), although it can run by itself, 
    it is more effective to run inside a container, so the two terms are used interchangeably. 
    \item A container bundles code and dependencies together, to provide reusability, reliability and time-efficient installation
    \item Containers also enable  technology-independent hyperautoamtion, as diverse technologies can be abstracted to work together
   \item Service definitions, versioning, auditing, retraining, maintenance, monitoring are further activities in AI operationalisation
      \item Data pipeline: availability, collection, storage, pre-processing, versioning, ethics
      \item AI pipeline: model compression, device compatibility, CI/CD
      \item Debugging AI: cross-functional complexity, silent errors
      \item Consider the turnaround time of a full AI development phase for major/minor revisions
\end{itemize}
\subsection{Hyperautomation processes and systems}
\begin{itemize} 
\item Hyperautomation (or intelligent automation) is the integration of AI capabilities with automated components in processes and systems
   \item Demonstrate capabilities of the AI solution to downstream/upstream process owners and stakeholders 
      \item Conduct a pilot phase of hyperautomation 
      \item Evaluate the outcomes of the process/workflow due to hyperautomation 
      \item Consider options for gains in efficiency (improve), effectiveness (enhance) or innovation (disrupt/transform)
\end{itemize}
\subsection{Monitor and evaluate performance}
\begin{itemize} 
   \item Evaluate model drift (decreasing accuracy), address this by re-training
   \item Evaluate model staleness (changing data/environment), address this by revising the model
   \item Evaluate end-user activity: adoption, questions, frequency of use, documentation, feedback
   \item Evaluate ROI from reduced costs (time, effort, skill)
   \item Evaluate ROI from increased revenue (new revenue streams, increased market share) 
   \item Evaluate ROI from others (reduced errors, increased productivity, reduced turnover)
\end{itemize}

\section{AI Algorithms to Applications}
Figure 2 depicts a mapping of algorithms with corresponding applications where the four primary AI capabilities of Prediction, Classification, Association and Optimisation are used as an intermediate layer. Although classification is a type of prediction,  we have separated the two, but detection is grouped along with classification. A further deliberation;
\begin{itemize} 
\item Prediction – regression, classification, time series, sequence
\item Classification (also detection) – object, anomaly, outlier concept 
\item Association – clustering, feature selection, dimensionality reduction
\item Optimisation – scheduling, planning, control, generation, simulation
\end{itemize}

\begin{figure} [H]
  \centering
  \includegraphics[width=1.0\textwidth]{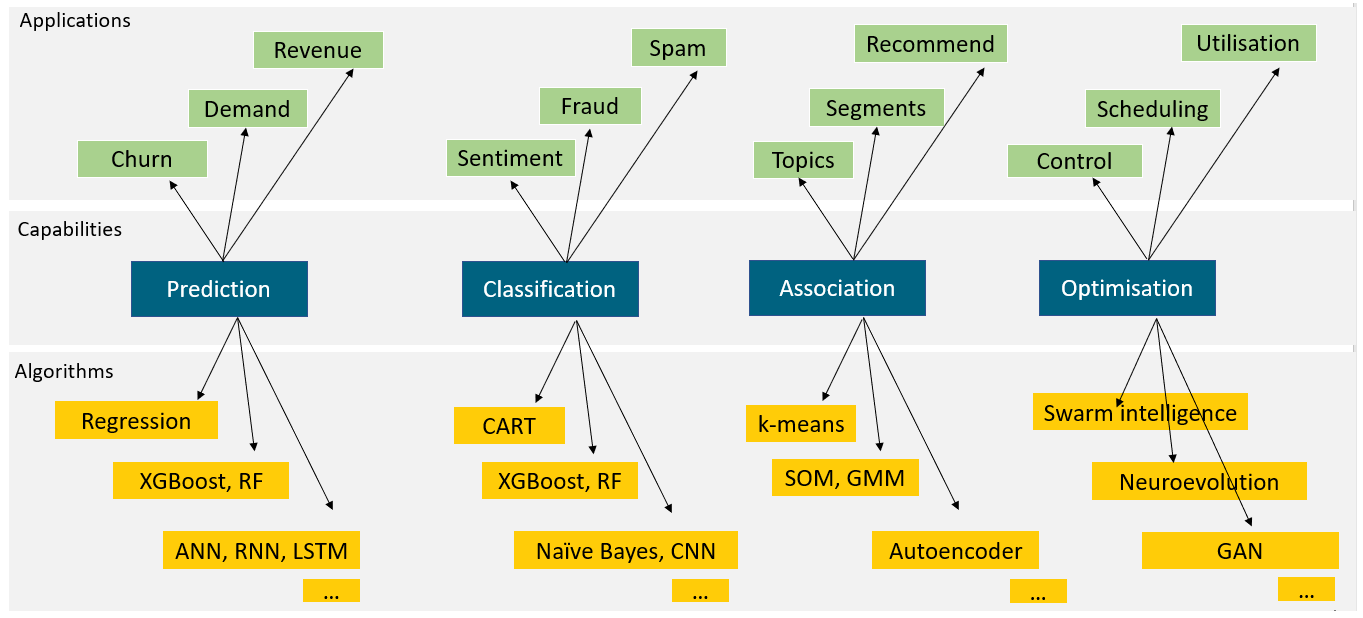}
   \caption{An Ontological Mapping of AI Algorithms to Capabilities and Applications}
  \label{fig:fig1}
\end{figure}

\section{AI Applications to Strategy}
And finally in Figure 3, we present an organisation context for this AI Life Cycle by positioning it within the flow of activities and information from organisational strategy to decision-making. The 17 stages have been condensed into five technical functions (in blue) to align with the organisational functions, depicted in gray. 

\begin{figure} [H]
  \centering
  \includegraphics[scale=0.5]{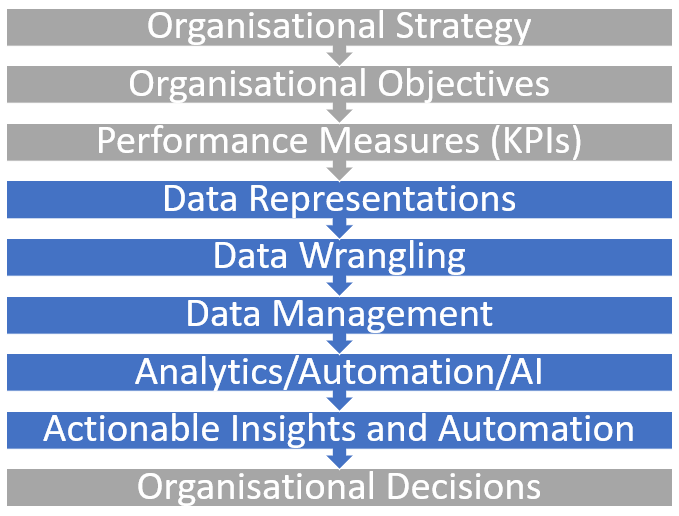}
   \caption{Organisational Context of the AI Life Cycle}
  \label{fig:fig1}
\end{figure}

\section{Conclusion}
In this brief article, we have presented the CDAC AI Life Cycle for the design, development and deployment of  AI systems and solutions. It consists of three phases, Design, Develop and Deploy, and 17 constituent stages across the three phases from conception to production. We anticipate the AI Life Cycle will contribute towards awareness, knowledge, and transparency of AI and its capabilities. The ontological mapping of AI algorithms to applications condenses all algorithms into four primary capabilities that will facilitate informed discussions between AI scientists, AI engineers and other stakeholders during the solution development process. The organisational context of the AI life cycle further integrates the AI team and their solutions with other stakeholders, such as  senior management and the executive in working towards an organisational strategy. 

%Bibliography
\bibliographystyle{ieeetr}  
\bibliography{references}

\end{document}